\documentstyle[12pt]{article}
\newcommand{\be}{\begin{eqnarray}}
\newcommand{\ee}{\end{eqnarray}}

\begin{document}

\rightline{RUB-TPII-12/98}

\begin{center}
{\large $N\to\Delta$ DVCS, exclusive
DIS processes and \\ skewed quark distributions in large $N_c$
limit}\\
\vspace{0.1cm}
{\bf
L.L.Frankfurt$^{a,b}$,
M.V. Polyakov$^{b,c}$, M. Strikman$^{b,d}$}\\
\vspace{0.1cm}
$^a$Physics Department, Tel Aviv University, Tel Aviv, Israel\\
$^b$Petersburg Nuclear Physics Institute, Gatchina, Russia\\
$^c$Institut f\"ur Theoretische Physik II, Ruhr--Universit\"at Bochum,\\
 D--44780 Bochum, Germany\\
$^d$Department of Physics, Penn State University\\
University Park, PA  16802, U.S.A.\
\end{center}

\begin{abstract}
\noindent
We evaluate the amplitude of Bethe-Heitler process:
$\gamma^* p\to \gamma \Delta^+$ -electron brems\-strahlung of
photon accompanied by the excitation of $\Delta$ isobar.  We show
that this background
is suppressed at small momentum transfer squared  $t$ to a proton
relative to that for $\gamma^* p\to \gamma p$
and $\gamma^* p\to \gamma \Delta^+$ - DVCS processes.
{}From experimental point of view, this means that $N\to N$ and
$N\to \Delta$ skewed quark distributions (SQD's) might be measurable
at small momentum transfer.
Several implications and applications of the QCD factorization
theorem for the processes $\gamma_{L}^*+p \to h_f +h_s$
are discussed where $h_f$ - the particle produced along photon momentum
$\vec q$ maybe either a meson or a baryon.
We discuss also $t$ dependence of DVCS and exclusive meson production
as the practical criteria to distinguish between soft and hard regime.
Basing on the large-$N_c$ picture of the nucleon as a soliton of
the effective chiral Lagrangian we  derive
relations between $N\to N$ and $N\to\Delta$ SQD's which can be used to
estimate amplitude of $N\to \Delta$ DVCS .

\end{abstract}

\underline{{\it Introduction}}

The aim of this presentation which summarizes talks given by the
authors at the workshop is
to
discuss feasibility to separate DVCS from
the background of inelastic processes related
to the
photon bremsstrahlung by electron and also
to discuss some applications of
the QCD factorization theorem for exclusive processes in DIS.

Recently, a new type of parton distributions
\cite{Bartels,AFS,Ji1,Rad,CFS}
has attracted considerable interest, the so called skewed (non-forward,
off--forward, non-diagonal) parton distributions (SPD's), which are
generalizations simultaneously of the usual parton distributions,
distribution amplitudes and the elastic nucleon form factors
in the case of vacuum  quantum numbers in $t$ channel. At the same time
there is a wide range of processes where one probes skewed densities
which do not have a diagonal analog\cite{CFS}.

The SPD's are not accessible in standard inclusive measurements. They can,
however, be measured in diffractive photoproduction of $Z$ boson
\cite{Bartels}, in diffractive electroproduction of vector mesons
\cite{AFS,CFS,Rad2,Rad3},in deeply--virtual Compton scattering (DVCS)
\cite{Ji1,Rad,Rad3} and in hard exclusive electroproduction
of mesons at moderate $x_{Bj}$ \cite{CFS,Piller}.

A quantitative description of these classes of processes requires not only
knowledge of the perturbative evolution of the SPD's,
but also non-perturbative information
in the form of the SPD's at some initial normalization point.
There were already model calculations of SPD's: in bag model \cite{Ji3}
and in chiral quark-soliton model \cite{pppbgw}. In the latter
calculation strong dependence of flavor singlet SPD's on skewedness
parameters was found. This dependence can considerably increase
DVCS amplitude at moderately small $x_{Bj}$ (see discussion below).

{}From experimental point of view the measurement of SPD's in DVCS is difficult
because of strong Bethe-Heitler (BH) background at small momentum transfer $t$.
To suppress the BH background one needs to increase the momentum transfer
and hence beam energy what makes problematic such measurements at TJNAF.
Alternative way to suppress the BH background is to study reaction
like $\gamma^* p\to \gamma \Delta^+$, in this reaction the BH background is
suppressed at small $t$ relative to that in $\gamma^* p\to \gamma p$
because in the former case the e.m. $N\to\Delta$ transition form factor is
always proportional to momentum transfer
as a consequence of conservation of e.m. current.
{}From theoretical point of view the measurements of $N\to\Delta$ SPD's
can give additional insight into the structure of the nucleon and $\Delta$.

\vspace{0.2cm}
\noindent
\underline{{\it $N\to\Delta$ BH process}}\\

The $N\to\Delta$ BH amplitude is expressed in terms of nucleon-$\Delta$
e.m. transition form factors. In our estimates we neglect
$G_{E2}^*$ and $G_C^*$ transition form factors relative to $G_M^*$
that, because the former numerically are very small
at small momentum transfer we are interested in. We compute the ratio
of amplitudes squared $R=|{\cal M}_{p\to \Delta^+}^{BH}|^2/|{\cal
M}_{p\to p}^{BH}|^2$ (suppression factor) in
the Bjorken limit. In a kinematic domain
$|t|\ll s$ . Here ``s'' is the square of of invariant energy of
$ep$  collision. One can
calculate leading term over powers of $s$ and then to use
conservation of em current to deduce  generalization of
the Weizs\"acker--Williams method of equivalent photon.
In this domain the suppression factor is given by simple formula:

\be
R=-\frac{\mu_{N\Delta}^2 (t-t_{min})}{4 M_N^2} \bigl(1+
\frac{t}{4 M_N^2}(\mu_p^2-1) \bigr)\; ,
\label{WW}
\ee
where $t_{min}$ is minimal  (in absolute value) momentum transfer
in a reaction $\gamma^* p\to\gamma \Delta^+$, which is
given at large $Q^2$ by:
\be
t_{min}=-\frac{x_{Bj}}{(1-x_{Bj})}\bigl(
x_{Bj} M_N^2 + M_\Delta^2-M_N^2\bigr)\; ,
\ee
$\mu_{N\Delta}=\sqrt{2/3}G_M^*(0)\approx (\mu_p-\mu_n)/\sqrt{2}$ is
transitional magnetic moment, $\mu_{p,n}$ are magnetic moments of
proton and neutron
in the Bohr magnetons.
{}From eq.~(\ref{WW}) we see that the $N\to\Delta$
BH process is suppressed at small momentum transfer, numerically at
$x_{Bj}=0.1$ and $t=-0.15$~GeV$^2$ the suppression factor eq.~(\ref{WW})
is about 0.18.

In the kinematic domain $x_{Bj}^2 M_N^2\sim (M_\Delta-M_N)^2\sim
|t|\ll M_N^2$ we derive the following expression for the
suppression factor:

\be
R=\mu_{N\Delta}^2
\frac{t-(M_\Delta-M_N)^2-t(t-t_{min})\frac{1-x_{Bj}}{2 x_{Bj}^2M_N^2}}
{\mu_p^2 t+4 (1-\frac{t}{4M_N^2}(\mu_p^2-1))(t+\frac{x_{Bj}^2
M_N^2}{1-x_{Bj}})\frac{1-x_{Bj}}{2 x_{Bj}^2}}\; .
\label{sec}
\ee
Evidently this expression is reduced to eq.~(\ref{WW}) when
$x_{Bj}^2 M_N^2 \ll |t|$.
Deriving eq.~(\ref{sec})
we assumed that the $N\to \Delta$
the shape of the transion form factor as well as the proton Sachs
form factors follow the dipole formula at small $t$.
Therefore in the ratio
$R$ the dipole factors cancel.
Numerically at $x_{Bj}=0.1$ and
$t=-0.1$~GeV$^2$ eq.~(\ref{sec}) gives
$R$ of about 1/5. Unfortunately, for the larger $x_{Bj}$ the
value of the factor $R$ for $t=t_{min}$ gets larger:
$R(x_{Bj}=0.1) \approx 0.15, R(x_{Bj}=0.2) \approx 0.35,
R(x_{Bj}=0.3) \approx 0.6$.
However, the $1/Q$ effects neglected here should be studied.

\vspace{0.2cm}
\noindent
\underline{{\it QCD factorization theorem for exclusive DIS processes -
some implications}}\\
\underline{{\it and applications}}\\
In \cite{CFS} the QCD factorization theorem was proven for the
process
\begin{equation}
     \gamma ^{*}(q) + p \to  M(q+\Delta ) + B'(p-\Delta )
\label{process}
\end{equation}
at large $Q^{2}$, with $t$ and $x=Q^{2}/2p\cdot q$ fixed.  It asserts that
the amplitude has the form
\begin{eqnarray}
   &&
   \sum _{i,j} \int _{0}^{1}dz  \int d\xi
   f_{i/p}(\xi ,\xi -x;t,\mu ) \,
   H_{ij}(Q^{2}\xi /x,Q^{2},z,\mu )
   \, \phi _{j}(z,\mu )
\nonumber\\
&&
   + \mbox{power-suppressed corrections} ,
\label{factorization}
\end{eqnarray}
where $f$ is an SPD,
 $\phi $ is the light-front wave function of the
meson, and $H$ is a hard-scattering coefficient, usefully
computable in powers of $\alpha _{s}(Q)$.

The proof of cancellation of the soft gluon interactions is
intimately related to the fact that the meson arises from a
quark-antiquark pair generated by the hard scattering.  Thus the
pair starts as a small-size configuration and only substantially
later grows to a normal hadronic size, in the meson.  This
implies that the parton density is a standard parton density
(apart from the skewed  nature of its definition).  For
example, no rescattering corrections are needed on a nuclear
target, other than those that are implicit in the definition of
universal parton densities, and that would equally appear in
ordinary inclusive deep-inelastic scattering. These statements
all apply to the leading power. This implies that the  theorem is valid
also for production of leading baryons
\begin{equation}
     \gamma ^{*}(q) + p \to  B(q+\Delta ) + M(p-\Delta )
\label{process1}
\end{equation}
and even leading antibaryons
\begin{equation}
     \gamma ^{*}(q) + p \to  \bar{B}(q+\Delta ) + B_2(p-\Delta )
\label{process2}
\end{equation}
Processes (\ref{process1},\ref{process2}) will provide a unique information
about multiparton correlations in nucleons.
For example, the process (\ref{process1}) will allow to
investigate what is probability for three quarks in a nucleon to
come close together without collapsing the wave function into
a three quark component
 - such probability would not be small
in meson cloud models of the nucleon and in say MIT bag model.
 On the other hand if one would try to follow analogy with
positronium, this probability would be strongly suppressed.
At the same time the (\ref{process2})
would allow probability (presumably numerically very small) to have
in a nucleon three antiquarks close together.

Note also that the process (\ref{process1})
could be used to produce a  $\rho$-meson with a zero momentum
if
\begin{equation}
q_0={Q^2+2m_{\rho}m_N-m_{\rho}^2 \over 2(m_N-m_{\rho})}.
\end{equation}
Hence in the case of nuclear targets it allows to
produce a $\rho$-meson with small momentum in the center of the nucleus.
Therefore process (\ref{process1}) could be used for
studing the medium modification of the properties of $\rho$-mesons.
Advantage of this process as compared to low-energy
processes is that effects of distortion of the  two pion
spectrum due threshold effects are less important in this case.


As non-perturbative input for QCD description of the process
(\ref{process1}) a new
mathematical
object (which can be called skewed
distribution amplitude (SDA)), beside usual DA for baryon $B$,
should be introduced. They are defined as
a non-diagonal matrix element of the tri-local quark operator
between meson $M$ and proton:

\be
\nonumber
&&\int \prod_{i=1}^3 dz_i^- \exp[i\sum_{i=1}^3 x_i\, (p\cdot z_i)]\,
\langle M(p-\Delta)|\varepsilon_{abc}
\, \psi^{a}_{j_1}(z_1) \,
\psi^{b}_{j_2}(z_2) \,
\psi^{c}_{j_3}(z_3)|N(p)\rangle\Bigl|_{z_i^+=z_i^\perp=0}\\
&&=\delta(1-\zeta-x_1-x_2-x_3)\, F_{j_1 j_2 j_3}(x_1,x_2,x_3,\zeta,t) \;,
\ee
where $a,b,c$ are color indices, $j_i$ are spin-flavor indices,
$F_{j_1 j_2 j_3}(x_1,x_2,x_3,\zeta,t)$ are new  SDA's
which depending on quantum numbers of meson $M$
can be decomposed into invariant spin-flavor structures.
These new
SDA's depend on variables $x_i$ (which are contracted with hard kernel in
the amplitude), on skewedness parameter $\zeta=1-\Delta^+/p^+$
(in some sense, with this definition of $\zeta$ the limit
$\zeta\to 0$ corresponds to usual distribution amplitude, $i.e.$
skewedness $\to 0$ means SDA $\to$ DA)
and momentum transfer squared $t=-\Delta^2$.


Though quantitative
calculations of  processes (\ref{process1}, \ref{process2})
 will take  time, some
qualitative predictions could be checked right away: the cross section
of the process for fixed $x$, and large $Q^2$
 should be proportional to the baryon
elastic form factor. In particular it would be instructive to study
the $Q^2$ dependence of the ratio of the the cross section of the
 process $\gamma ^{*}(q) + p \to p +
\pi^0$ and the square of the elastic proton form factor.
If the color transparency suppresses the final state interaction
between the fast moving nucleon and the residual meson state early enough
one may expect that this ratio may reach the scaling limit in the region where
higher twist contributions to the nucleon form factor are still large
(It is worth emphasizing that the longitudinal
distances involved in the final state interaction of the system flying
along $\vec q$ with the residual system are
 much smaller in this case than in the case of
$A(e,e'p)$ reaction, so the expansion
effects would be much less important in this case.
Another  interesting process is $\gamma^{*}(q) + p \to \Delta^{++} +
\pi^-$  may allow to compare the wave functions of $\Delta$-isobar
and a nucleon in the way
complementary to the $N\to \Delta$ transition processes.

Let us mention also two other applications of the factorization theorem:

(i) If
in
the process (\ref{process}) the leading meson is exotic - either
a $q\bar q g$ or $q\bar qq\bar q$ state, the cross section of this process
should decrease with $Q^2$ much faster than in the case of
 the $q\bar q$ mesons - by $Q^4$ in the  $q\bar q g$ case, and
 $Q^8$ in the  $q\bar q q\bar q$ case.
Due to a mixing with $q\bar q$ states this fast drop of the cross section
with $Q^2$ would be followed by a slower decrease but with much smaller
absolute cross section than for the $q\bar q$ mesons.
Thus the study of the $Q^2$ dependence of the production of candidates to
the exotic mesons could help to check their quark-gluon content.

(ii) Study of the meson spectrum in the process $\gamma^* +p \to p +M$
may help to investigate the role of pions in the nucleon wave function.
In a naive model of the nucleon with
a pion cloud one would expect that production of a pion will dominate.
However if more complicated nonlinear pion  fields are important in
the low $Q^2$ $q \bar q$ sea, one may expect that
 higher
recoil
masses
 would be at least as important.
One can go a one step further and ask whether if one takes the valence
three quarks out of the nucleon - could
 the  residual system couple
strongly enough to the gluonium states, providing another avenue for
looking for exotic meson states.

\vspace{0.2cm}
\noindent
\underline{{\it How to distinguish experimentally
between hard and soft QCD regimes?}}\\

Here we want to explain important advantage of identification of hard
QCD physics in DVCS and more generally in hard exclusive processes as
compared to that in e.m. form factors of hadrons. In both processes
there is competition between soft QCD physics (end
point contribution=Feynman mechanism)
and hard QCD physics
\cite{BFGMS,Radyushkinthisconf}.
Serious problem for
the investigation of  e.m. form factors of hadrons
is absence of
 unambiguous
criteria
to distinguish between
both contributions \cite{Stolerthisconf}.
On the contrary the dependence of hard exclusive processes on momentum
transfered to proton can be used as
an  unambiguous
criterion
 of
the
dominance
of soft or hard physics in the process.

Cross section of
a
 two body process can be parameterized at large energies as
$d\sigma(\gamma^* + p \to M +T)/dt= A\exp(Bt)$. Here $M$ can
be photon,vector or pseudo-scalar meson. The slope $B$ can be parameterized as
$B=B_0(Q^2)+2\alpha'(Q^2)\ln(s/\mu^2)$. Common wisdom based on the
success of Regge pole hypothesis in the description various two body
high energy processes is that for the soft QCD processes $B$ should be
independent on $Q$ but it should depend on the energy $s$. We
presented above the parameterization of cross section which is
valid for the exchange by Regge pole.
On the contrary for hard exclusive processes scale of dependence of
hard vertex on $t$ is given by $Q$. Thus $B_0 \rightarrow B_N$
in the limit of large $Q$.
Here $B >B_N$  and $B_N$ should be the same for all hard processes
with the same quantum numbers in the crossed channel
\cite{AFS}. Moreover $\alpha'(Q^2)\rightarrow 0$ also in the limit of
large $Q$. This is because QCD evolution for
skewed parton distributions prevents Gribov diffusion of small
gluon wave package to large impact parameters.
Predicted dependence of $B$ for the hard exclusive processes on $Q^2$
agrees with HERA data on diffractive production of
vector mesons for large $Q^2$ \cite{hera}.
     The difference between soft and hard processes should be
most striking for the processes with non-vacuum quantum numbers in the
crossed channel like excitation of $\Delta$. Really
in the  case of
the
usual Regge pole trajectories; $\alpha'
\approx
 1 GeV^{-2}$
and dominance of
the Regge pole exchange
usually  reveals
 itself  at the projectile energies
of few GeV - famous Dolen-Horn-Schmidt duality. Thus
the decrease of the slope $B$ may provide
an evidence for
the dominance of hard QCD is
with increase of $Q^2$ already
at moderately large $s$.
So this physics could be studied already at TJNAF and HERMES energies
(physics for TJNAF, HERMES and COMPASS). The limiting value of
$B(Q\rightarrow \infty)$ should be significantly smaller than for $Q=0$
and independent of energy and $Q^2$.

\vspace{0.2cm}
\noindent
\underline{{\it $N\to\Delta$ skewed quark distributions}}\\
The soft part of the DVCS $N\to\Delta$ amplitude
is parameterized generically  in terms of eight ($\times N_f$)
$N\to\Delta$ skewed quark distributions (SQD's)
($cf.$ four($\times N_f$) for $N\to N$ SQD's) [the detailed
expression will be given elsewhere \cite{elsewhere}].

In order to relate $N\to\Delta$ SQD's to $N\to N$ ones we shall use the
large $N_c$ picture of the baryons. In this picture the nucleon and
$\Delta$ are different rotational states of the same
object--the ``classical" or ``generalized" nucleon. Therefore the
static properties of nucleon and $\Delta$ can be related to each other,
a number of such relations were derived in the past (see
\cite{soliton}) and it turns out that such relations work very well.
Let us give a few examples of such relations:

\be
\frac{\mu_{N\Delta}}{\mu_p-\mu_n}=\frac{1}{\sqrt 2}\;
({\rm expt.}\; 0.71\pm 0.01),\quad \frac{g_{\pi N \Delta}}{g_{\pi N N}}=
\frac{3}{2}\; ({\rm expt.}\; 1.5\pm 0.12).
\ee
Such kind of relations is independent of particular
dynamical realization of idea baryon as chiral soliton.

Below we sketch an idea how to relate $N\to \Delta$ SQD's to
$N\to N$ ones, detailed account of the calculations will be given
elsewhere \cite{elsewhere}.
First we  write the matrix element of bilocal quark operator
on the light cone between soliton (``generalized" nucleon) states.
[Technique how to calculate such matrix elements was developed
in \cite{DPPPW}, we refer the reader to this paper for details]
These
matrix elements are expressed in terms of operators in collective
coordinate space.  The projection on a baryon state with given spin and
isospin components is obtained by integrating over all spin-isospin
rotations, $R$ \cite{soliton},
\be \langle S'=T',S'_3,T'_3|\ldots|
S=T,S_3,T_3\rangle &=& \int dR\;\phi^{\ast\;S'=T'}_{S'_3T'_3}(R) \; \ldots
\; \phi^{S=T}_{S_3T_3}(R)\,.  \label{spisosp} \ee
Here
$\phi^{S=T}_{S_3T_3}(R)$ is the rotational wave function of the baryon
given by the Wigner finite-rotation matrix \cite{soliton}:
\be
\phi _{S_3T_3}^{S=T}(R) &=& \sqrt{2S+1}(-1)^{T+T_3}D_{-T_3,S_3}^{S=T}(R).
\label{Wigner}
\ee
Analogously, the projection on a baryon state with given momentum
${\bf P}$ is obtained by integrating over all shifts, ${\bf X}$, of the
soliton,
\be
\langle {\bf P^\prime}|\ldots|\ {\bf P}\rangle
&=& \int d^3{\bf X}\;e^{i({\bf P^\prime-P})\cdot{\bf X}}\; \ldots
\label{totmom}
\ee

It can be shown \cite{elsewhere} that in the leading order of
$1/N_c$ expansion the soliton (the ``classical" or``generalized"
nucleon) is characterized (for $N_f=2$) by 4 SQD's
(7 in the next to leading order).
Because both $N\to N$ and $N\to\Delta$ SQD's are expressible in terms
of the $same$ SQD's of the ``generalized" nucleon there are a number of
relations between them. Because of lack of space we give here the
simplest relation between DVCS amplitudes squared for $N\to  N$ and
$N\to\Delta$ transitions in the leading order of $1/N_c$ expansion:

\be
\biggl|{\cal M}^{DVCS}_{p\to\Delta^+}\biggr|^2=\frac 12
\biggl|{\cal M}^{DVCS}_{p\to p}-{\cal M}^{DVCS}_{n\to n}\biggr|^2
\bigl(1+O(\frac{1}{N_c})\bigr) \; .
\label{glav}
\ee
A few comments are in order here. First, the above relation is hold
for Bjorken $x_{Bj}$ of order of $1/N_c$, $i.e.$ in valence region,
it is modified at $x_{Bj}$ close to unity (see discussion in
\cite{DPPPW}). Second, the $1/N_c$ corrections (denoted
$O(1/N_c)$ in eq.~(\ref{glav})) are also expressible solely in terms of
$N\to N$ SQD's and hence can be estimated numerically, the
corresponding results are in preparation \cite{elsewhere}.

The simplest estimate of $N\to\Delta$ DVCS amplitude squared
eq.~(\ref{glav}) shows that the $N\to\Delta$ DVCS amplitude
is comparable with $N\to N$ that. The detailed estimate will
be given elsewhere \cite{elsewhere}.

\vspace{0.2cm}
\noindent
\underline{{\it Concluding remarks}}\\

We obtained that the BH background to the $N\to\Delta$ DVCS
is suppressed at small momentum transfer, the BH cross section
behaves $\sim t^0$ at small momentum transfer, whereas cross section of
the $N\to N$ BH process behaves like $\sim 1/t$.
The DVCS cross sections in both cases are of the same order.
This makes the measurements of the $N\to\Delta$ DVCS favorable at
(upgraded) TJNAF energies. Such measurements can give new information of
nucleon and $\Delta$ inner structures. Interesting to note that
$N\to\Delta$ DVCS at small momentum transfer is especially sensitive to
the helicity dependent SPD's.

Another interesting possibility is to measure the reaction
$\gamma^* N\to\gamma (k)\pi N$ ($k=1,2,\ldots$) with invariant mass of
the $(k)\pi N$ system below and in resonance region.
It is correct that BH background in this case is suppressed at small $t$.
The process $\gamma^* N\to\gamma (k)\pi N$ is
described by the generalized SPD's, which additionally to $x$, skewedness
and $t$ depend on invariant mass of $(k)\pi N$ system and on the
distribution of longitudinal momentum between the final hadrons. These
generalized SPD's can be calculated in chiral quark soliton model.
Anyway, study of say $N\to \pi N$ SPD's is required to estimate
non-resonant background to $N\to\Delta$ DVCS\footnote{M.P. is grateful
to C. Hyde-Wright for discussion of this point}.

Let us also note that the estimates of the DVCS cross section made in
\cite{guichon} were based on oversimplified models of SPD's. In
particular, the dependence of the SPD's on skewedness parameter was
neglected and the SPD's which correspond to helicity flip amplitudes
were neglected. The authors of ref.~\cite{guichon} correctly
argue that the  contribution of the latter SQD's is small at small $t$.
But contribution of the pion pole of
the type $\sim g_A^2 t/(|t|+m_\pi^2)^2$ may appear
not small at $|t|\sim m_\pi^2$.

In our opinion the simulations of the DVCS with more
realistic models for SPD's are needed.

Also one should try to design experiments which would
be able to study a  broad range of processes
(\ref{process},\ref{process1},\ref{process2})
including channels with strangeness. This would allow to expand
immensely our understanding of the short-range hadron structure.

\end{document}